\documentstyle[12pt]{article}

\def\be{\begin{eqnarray}}
\def\ee{\end{eqnarray}}

\def\d{\delta}
\def\D{\Delta}

\def\Ho{\overline H}

\def\G{\Gamma}
\def\Gt{\tilde G}

\def\f{\varphi}
\def\l{\lambda}
\def\o{\omega}
\def\O{\Omega}

\def\cg{c^{\dagger}}

\def\dg{\dagger}

\def\kt{\tilde k}
\def\s{\sigma}

\def\ur{\uparrow}
\def\dr{\downarrow}

\def\cg{c^{\dagger}}
\def\e{\epsilon}

\begin{document}
\vspace{0.4in}
\begin{center}

{\bf Influence of spin structures and nesting on Fermi surface and
a pseudogap anisotropy in t-t'-U Hubbard model.}  \\

\vspace{0.4in}
{\bf \fbox{A.A.Ovchinnikov} and M.Ya.Ovchinnikova.}   \\
\vspace{0.2in}
{\it Institute of Chemical Physics, RAS,
Moscow, 119977,  Kosygin str., 4.} \\
\vspace{0.85in}
\end{center}

%
%
%
\begin{abstract}
Influence of two types of spin structures on the form of the Fermi surface
(FS) and a photoemission intensity map is studied for $t-t'-U$ Hubbard model.
Mean field calculations are done for the stripe phase and for the spiral
spin structure. It is shown, that unlike a case of electron doping, the
hole-doped models are unstable with respect to formation of such structures.
The pseudogap anisotropies are different for $h-$ and $e-$ doping.
In accordance  with ARPES data for $La_{2-x}Sr_xCuO_4$ the
stripe phases are characterized by quasi-one-dimensional segments of FS in
the vicinity of points $M(\pm \pi,0)$ and by suppression of spectral weight
in diagonal direction $k_x=k_y$. It is shown that spiral structures display
the polarisation anisotropy: different segments of FS correspond to
electrons with different spin polarisation.

\end{abstract}

PACS: 71.10.Fd,  74.20.Rp, 74.20.-z

The angular resolved photoemission spectroscopy (ARPES) is effective in the
study of electronic structure of cuprates \cite{1,2}. It gives the image
of the Fermi surface (FS) projection onto a $CuO_2$ plane. Conclusions of
the early works (see \cite{1,2}, and references therein)
were consistent with FS of the hole
type centered at point $Y(\pi,\pi )$ of two-dimensional Brillouin zone (BZ).
Later other versions of the FS topology have been discussed. In particular,
a presence of an electron type of the FS have been proposed with the center
at $\G (0,0)$ in $Bi_2Sr_2CaCu_2O_{8+\d}$ (BSCCO) \cite{3}.
Revision of the problem \cite{4,5} partly confirms the original version. At
the same time for $La_{2-x}Sr_xCuO_2$ (LSCO) it has been proved a transition
from  h-type of the FS to  e-type during
transition from underdoped (UD) to overdoped (OD) regions of the phase
diagram \cite{6,7}.
The ARPES data gave an evidence of the d-wave superconducting gap and
of the opening of the pseudogap (PG) in UD BSCCO. Recently a
bilayer splitting of bands
\cite{8,9,10,11} in BSCCO and the time reversal symmetry breaking in
UD BSCCO \cite{12} have been discovered.

A wide use of the photoemission intensity maps in the space $k_x, k_y, \o$
poses a problem of extracting the FS from the ARPES data. One aspect of
the problem, calculation of the matrix elements is discussed in
\cite{13,14}. However the FS topology and intensities of various (main
or shadow) segments of the FS depend also on the spin and charge structures
of a ground state of a system.

The goal of the present work is a model study of the influence of periodic
structures
on the FS and the photoemission intensity based on the t-t'-U Hubbard
model. Unlike the static structures in manganites, in cuprates one deals with
rather dynamical structures with a time scale larger than time
$t>10^{-6}\div 10^{-9}$ sec in the $\mu$-SR experiments. So at small
times or for the processes with energy resolution higher than $\hbar/t$
the local spin structures may be considered in quasi-static limit.
Then the question arises about a correspondence of the structures to
the ARPES data. The experimental indications of the SDW and CDW structures
in cuprates are the incommensurate peaks in the spin susceptibility at
$q=(\pi\pm\d q,\pi), (\pi,\pi\pm\d q)$ in LSCO \cite{15},
linear structures along the $CuO_2$ bonds with period 4$a$ ($a$ is a
lattice constant) observed in the tunnel spectra
\cite{16}, periodic chess structure $4a\times 4a$  around the vortex
in mixed state of BSCCO \cite{17} etc.

Here we  study the stripe and spiral structures in the
simplest mean field (MF) approach,
interpret the PG, discuss examples  of structures with
the different types of the PG anisotropy and to connect some properties
of MF solutions with  features observed in some cuprates. Earlier
\cite{18} the DDW (d- density wave) structure has been proposed as a
possible hidden order parameter (OP) in cuprates.  Here a search
of possible OP
is extended to stripe and spiral structures. Stabilization of such
structures originates from the removing of a degeneracy
of states in "hot points"
- the van Hove singularities (VHS) or in parallel segments of the FS at
the so-called nesting.

Consider a t-t'-U Hubbard model with the zero band energies
\be
\e_k=2t(\cos{k_x}+\cos{k_y}) +4t'\cos{k_x}\cos{k_y}
\label{1}
\ee
Further we put $t=1$, so all energies are measured in units of $t$.
In MF approximation this Hamiltonian is insufficient to describe
the superconducting (SC) pairing. It may be supplemented by
an empirical pairing interaction of electrons
at neighboring centers deduced in more accurate approaches (for example,
\cite{19}). But here we study a normal state and therefore retain
the original $t-t'-U$ model.

A periodic 2D structure is determined by  two translation vectors
\be
{\it E}_{1}=({\it E}_{1x},{\it E}_{1y}), ~~
{\it E}_2=({\it E}_{2x},{\it E}_{2y})
\label{2}
\ee
and the corresponding vectors of the inverse lattice $B_1,B_2$
which satisfy the relations
${\sl E}_i B_j=2\pi \d_{ij}$. (The components of ${\sl E}_i$ и $B_i$
are in units of the corresponding constants of the original
and inverse lattices).
The unit cell of the structure contains $n_c$ sites with coordinates
$j=(j_x,j_y)$, so that each site
$n=n(L,j)=(n_1 ,n_2 )={\sl E}_1 L_1+{\sl E}_2 L_2+(j_x,j_y )$
is determined by the integers $L_1 , L_2$
determinig the position of the unit cell and
by integers $j=(j_x,j_y)$, fixing the site inside the unit cell.

Let $\kt$ is the quasi-momentum inside the main BZ ${\tilde G}$ of
a given periodic structure unlike
the momentum $k$ from the BZ $G$ of the original lattice. The areas
of $\Gt$ and $G$ are restricted by conditions $|\kt B_i|\leq\pi$ or
$|k_{x(y)}|\leq\pi$. The order parameters of the periodic structures are
the mean charges and
spin densities on sites of the unit cell
\be
r_j={1 \over N_L}\sum_L<r_{n(L,j)}>;
~~S_{\mu j}={1 \over N_L}\sum_L<S_{\mu,n(L,j)}>
\label{3}
\ee
Here $\mu$ numerates the spin components, $N_L=N/n_c$ is a number of the unit
cells,  $n_c$ is a number of sites in it.
In the MF approximation the mean energy (1) is
$
\Ho =<T>+N_L U\sum_j(r_j^2-\sum_{\mu}S_{j\mu}^2)
$
and the wave function is determined by the occupation of the one-electron
eigenstates $\chi_{k\l}^{\dg}$  of the linearized  Hamiltonian
$$H_{lin}=T+N_L\sum_j\{2Ur_j {\hat r}_j-2US_{\mu j}{\hat S}_{\mu j}\}
=\sum_{{\kt}\in\Gt}{\hat h}_{\kt}.$$
The latter is divided into independent contributions from each
reduced quasi-momentum $\kt$. Here ${\hat r}_j,{\hat S}_{\mu j}$ are the
operators corresponding to averages (3).
In the momentum represen\-ta\-tion the eigenstates of ${\hat h}_{\kt}$
are expanded over a
basis set of $2n_c$ Fermi operators
\be
\chi_{{\kt}\l}^{\dg}=\sum_{m,\s}\cg_{\kt+Bm,\s}W_{m\s,\l}(\kt ),
\label{4}
\ee
where $\l=1,\dots,2n_c$, $Bm=B_1 m_1+B_2 m_2$  and
$m=(m_1,m_2)$ is such set of integers, that allows the vectors ${\kt+Bm}$
to share all phase space $G$.

The matrix of eigenvectors $W_{m\s,\l}$ and  the eigenvalues $E_{\kt,\l}$
are determined by diagona\-li\-za\-tion of $h_{\kt}$  in the basis set
$\{\cg_{\kt+Bm,\s}\}$:
\be
(h_{\kt})_{m\s,m',\s'}W_{m'\s',\l}=W_{m\s,\l}E_{\kt ,\l}
\label{5}
\ee
Here
\be
(h_{\kt})_{m\s,m',\s'}=\d_{m m'}\d_{\s\s'}\e_{\kt+Bm}+
U\sum_j \f(j,m'-m)[r_j\d_{\s\s'}-S_{\mu j}(\s_{\mu})_{\s\s'}]
\label{6}
\ee
with
$
\f(j,m)=\exp{[i(j-j_0)Bm]};~~ Bm=B_1 m_1+B_2 m_2.
$
In turn, the order parameters (3) are expressed as
\be
\{r_j , S_{\mu j}\}={1 \over {2N}}\sum_{\kt\in\Gt}\sum_{ms,m's'}
\{\s_0, \s_{\mu}\}_{s s'}\f(j,m'-m)
W^*_{ms,\l}(\kt)W_{m's',\l}(\kt)f(E_{\kt \l}-\mu)
\label{7}
\ee
Here the Puali matrices $\s_{\mu}, ~\s_0$ in (9) correspond
to the components $S_{\mu j}, r_j $
and  $f$ is the Fermi function. Equations (9) determine
the self-consistent MF solution.

The intensity of the photoemission of electron with a projection $k$
of the momentum on $ab$ plane and the energy $E=h\nu-\o$ is
\be
I(k,\o)=|M(k)|^2 A(k\o)f(\o)\otimes R_{\o k}.
\label{8}
\ee
It is determined by the matrix element $M(k)$,
a spectral density $A(k\o)$ and the Fermi function $f$.
Usual convolution is done with the Gaussian function
$R_{\o k }$ \cite{13} with parameters, which characterize  a finite
resolution over $k$ and energy.
The dependence of the matrix element $M$ on $k$ was studied in
\cite{13,14}. Here for simplicity we take a constant value for it,
since we study an effect of the structure on the spectral density $A$.
In the one-electron approximation
\be
A(k,\o)={1 \over N} \sum_{\kt\in\Gt}\sum_{m,\s,\l}
|W_{m\s,\l}(\kt)|^2{\overline \d} (E_{\kt \l}-\mu-\o) \d_{k,\kt+Bm}
\label{9}
\ee
Here $\l=1,\dots , 2n_c$ and index $m=(m_1 , m_2 )$ numerates all
independent Umklapp vectors $Bm=B_1 m_2 +B_2 m_2$.
A standard replacement of the $\d$- function in (9) by a
function with finite
width $\O$ is implied. A map of $I(k_x,k_y, \o=0)$ allows to visualize both
the main and shadow segments of the FS. Though the band energies are
periodic functions in $k$ space ( $E_{\kt+Bm,\l}=E_{\kt,\l}$
for any $m=(m_1,m_2)$ ), the intensity (8) does not posses such periodicity.
Therefore even for the constant matrix element $M$ in (8) various segments
of the FS display themselves with a different intensity due to the Umklapp
processes in the case of SDW or CDW structure.

The intensity map allows easily to repeat the known results
\cite{20,21,22,23,24,25,26} about changing of the FS topology with
increase of the doping and opening the pseudogap in UD region for the
homogeneous AF solutions. The energies of the lower and upper Hubbard
bands are
\be
\e_{\kt\l}=4t'\cos{k_x}\cos{k_y}\pm \sqrt{U^2 d_0^2
+4t^2(\cos{k_x}+\cos{k_y})^2}+const
\label{10}
\ee
The VHS in the DOS of the lower band correspond to the "hot points"
$M=(\pm\pi,0),(0,\pm\pi)$.  A form of the FS critically depends on a sign
of $t'$. At $t'=0$ energy $E_\l (k)$ is constant for all $k$ along the
magnetic Brillouin zone (MBZ) boundary. At $t'>0$ the energies at the points
M are lower than at the diagonal points. This leads to
the formation of the known
hole pockets around points $(\pi/2,\pi/2)$ or the electron pockets
around $(\pi,0)$, $(0,\pi)$ in UD case of the h- or e- doped models
\cite{21}. In the same UD region the PG is open
at $x,y$ directions or at diagonal directions for h- or e- doped models.
Such a topology is determined by a profile of the energies $E_{\l}(k)$ of
lower or upper Hubbard bands as a function of $k$ on the MBZ boundary.
But this profiles are greatly  influenced also by a formation of the spin or
charge structures in system.

The MF calculations allow to test a stability or an instability
of the homogeneous AF states with respect to the
formation of the periodic spin or charge structures.
The Fig.1 presents the mean energy $\Ho =<H>$ as a function of doping
$\d=|1-n|$
for the normal state of the hole- and electron- doped models for a set of
structures. In addition to homogeneous paramagnet (PM) and AF
solutions the next structures are considered:

1) The stripe structure which consists of the antiphase AF stripes parallel
to the y-axis. A fixed width 4a of the stripe (8 sites in a unit cell
of the structure) is taken and
the domain walls are chosen as centered on bonds or on the sites. Both
structures have a similar form of the FS, but that
with the bond-centered domain walls occurs to have some
lower energy and only it is discussed below.

2) The spiral spin structures
\be
<S_n>=d_0 [{\bf e}_x \cos{Qn} +  {\bf e}_y \sin{Qn} ]
\label{11}
\ee
with spirality vector $Q=Q_x=\pi (\eta,1)$ or $Q=Q_{xy}=\pi(\eta,\eta)$,
with $\eta\sim 0.75 - 0.8$.

3) The checkered structure with antiphase AF square domain $4a\times 4a$.
It have a higher energy than the  above two structures, and unrealistic
shape of the FS, so the results are not presented here.

4)  The proposed in \cite{18} states of the orbital antiferromagnet
with staggered currents on a plackets. In MF approximation they
do not appear for $t-t'-U$ Hubbard model without additional
interactions.  At hole doping and $U/t\ge 5$ there exist the MF solutions
with staggered spin currents and rotations on $\pi/2$ of local spins of the
neighbor sites. But their energies are higher than those of stripe and
spiral structures and corresponding results are not presented here.

Most  of calculations were carried out for structures with fixed commensurate
period $8a$, though the optimal size of the AF domains or spiral vector
depend on doping. For spiral states the dependence of $Q(\d)$ has been
calculated  many times. Here we are interested only in the specific
features of the FS and the pseudogap anisotropy.

Fig.~2 shows that at the h-doping the homogeneous AF solutions are
unstable with respect to formation of the stripe and spiral structures.
Formation of these structures extend the doping range where the local spin
retains a nonzero value (a disappearing of the $<S_{\mu j}>$ corresponds to
matching of the energy of some structure and that of the paramagnet state).
At the same time at e-doping the MF energies of the above structures
appear to be higher than the energy of the homogeneous AF solution.
Large stability of the AF state in the e-doped models corresponds
to the more wide region of the AF order in cuprates of the n-type
than in ones of the p-type. The result about the absence of the stripe
and spiral structures in the n-type models  are consistent also with
a commensurate peak at  $Q\sim (\pi,\pi)$ in the spin susceptibility
observed in NCCO, PCCO (unlike the incommensurate peaks in the hole doped
cuprates).

Fig.2a presents a map of the intensity  $I(k,\o=0)$, Eqs.(8,9),
for the stripe structure
in the model with $U/t=6$, $t'/t=0.1$ at $n=0.8$.
In case of AF MF solution the chosen doping is close to the "optimal"
doping at which the Fermi level passes through the VHS in the DOS of the
lower Hubbard band. The stripes (which are parallel to an y-axis)
split the VHS.  The "hot points" $M_x=(\pi,0)$ and $M_y=(0,\pi)$
become non-equivalent points and the quasi-1D asymmetric
horizontal segments of the FS are formed. Fig.2b presents the intensity map
symmetrized over the structures with the x- or y- stripe orientation.
The FS form differs strongly from that for the homogeneous AF
state of the model with $t'>0$. The main difference is the absence of
the FS in the diagonal direction and near the $M_y$. This gives an evidence
that the stripes open the pseudogap in these regions of $k$.
The corresponding band
energies $E_{\l}(k)$ descend below the Fermi level and the
work function $\D_{PG}=\mu-E_{\l}>0$ for the electron to be removed
from such $k$ is observed as the PG in ARPES data.
The revealed anisotropy of the FS and pseudogap may change our conclusions
about possible symmetry of the SC order in the striped state.
The retaining of the quasi-1D segments of the FS only suggests the
1D character of conductivity along the stripe direction (here y-axis).
With the increase of $t'$ up to the value 0.3 there appear
the small additional hole pockets at  around $k_S=(\pi/2,\pi/2)$
in the intensity map.

The anisotropy of the FS and the PG suggests the revision
of the possible symmetries of SC order coexisting with the stripes.
If the structure is symmetric with respect to replacement
$x\leftrightarrow y$, then the
d-wave SC order $<\cg_{k\ur}\cg_{-k\dr}>\sim \cos{k_x}-\cos{k_y}$
is expected. It
provides the orthogonality of the pair function to the on-site pair
function $\cg_{n\ur}\cg_{n\dr}$ which is suppressed by the on-site
repulsion U. Taking into account a new FS view (see Fib.2a) one may
suppose the extended s-wave SC pair function
$<\cg_{k\ur}\cg_{-k,\dr}>\sim \cos{k_x}+\cos{k_y}$.  The latter
might be orthogonal to the on-site pair function due to the node lines
$k_x\pm k_y=\pm \pi$. Unlike the gain of the node lines in the angular
dependence of
the d-SC gap, one  can suppose a gain of the node in an analogue of a
"radial" part of the pair function. Verification of such hypothesis requires
further calculations.

Fig.3a presents the one-electron energies $E_{\l}(k)$ (eigenvalues of MF
problem) as functions of quasi-momentum varying along the contour
$$Y(-\pi,\pi)-M_y(0,\pi)-Y(\pi,\pi)-M_x(\pi,0)-Y(\pi,-\pi).$$
As expected, band energies $E_{\l}(k)$ are periodic functions with
the period $\pi/4$ on the first horizontal segment of contour
$Y-M_y-Y$.  However, the intensity map selects only the
energy levels which are non-shadow  at given  $k$. Fig.3b presents such
map of $I(k,\o)$  on the plane $k,\o$ for $k$ varying along the same
contour.  Intersections of band energies with the Fermi level in the vicinity
of $M_x$ correspond to quasi-1D FS of Fig.2a.

The above pictures may be explained by an action of the spin-dependent
average field. The main harmonic of this field is
$F(n)\sim \cos{\{Q_{\eta}n+\f_0\}}$
with $Q_{\eta}=(\eta\pi,\pi)$ (here $\eta=0.75$).
This field shifts upstairs the zero level $\e(0,\pi)$ at $M_y$
pushing it from the lower levels $\e_k$ at $k=(\pm \eta\pi,0)$
near $M_x$. The same field descends downstairs (lower the Fermi level)
the zero level $\e(\pi,0)$ at $M_x$ pushing it from
the higher zero levels $\e(\pm 0.25\pi,\pi)$ near $M_y$.

The ARPES data for UD LSCO \cite{6,7,27}  are in qualitative agreement
with the properties of the model calculations. It was shown \cite{6,7,27}
a presence of two different segments of the FS - near the points $M$
and at diagonal directions, systematic suppression of spectral weight near
$(\pi/2,\pi/2)$ in comparison with BSCCO or OD LSCO, the straight segments
of the FS near $(\pi,0)$ with the width $\sim \pi/2$. All these
observed features have been interpreted as an evidence of nonhomogeneous
stripe structures in UD LSCO \cite{6,7,27} or the combine action of the
stripe structure and of the order-disorder \cite{27a}.
Our calculations confirm this interpretation.
Another evidence of the stripe structure is given by the observation
of the "incommensurate" peaks in neutron scattering at $k=(\pi\pm\d, \pi)$,
$(\pi, \pi\pm\d)$ in LSCO \cite{15}.

Another type of the spin and angular anisotropy of the FS manifests
itself in spiral spin structures (11)
which possess the polarisational anisotropy of the FS.
Different segments of the FS correspond to the electrons with different
preferential spin polarisation.  Really,  the average field from spiral
spin structure mixes one-electron states $\{\cg_{k,\ur},\cg_{k+Q,\dr}\}$
provides a splitting of the VHS's. But unlike the stripe structures,
the occupations $n_{k,\s}=<\cg_{k\s}c_{k\s}>$
and intensities $I_\s (k\o)$ of the photoemission of the electrons with
fixed polarisation $\s$  depends on this polarisation $\s$.
Values $I_\s (k\o)$ are determined by Eqs.(8,9), but without
summing over $\s$ in right part of (9).

Fig.4a presents a map of the intensity $I_{\s=\ur}(k,\o=0)$ for the spiral
state with $Q=(0.8\pi,\pi)$ and for the spin polarisation $\s=\ur$
on the axis $z$ normal to the plane of the spin rotation of the spiral
structure (11). Thus a photoemission of electrons with the up spin
polarization ($\s=\ur$) corresponds to the strongly anisotropic FS.
For the opposite spin polarisation the FS coincides with the FS of
Fig.3a, reflected in the plane $x=0$. Fig.3b presents the FS, symmetrized
over the spins and over two types of structures with
$Q=(0.8\pi,\pi)$ и $Q=(\pi,0.8\pi)$. In the vicinity of the point $M$ the FS
have both the intersection with the line $M-Y$ typical for the FS of the
h-type and the same with the line $M-\G$ typical for the e-type of the FS.
Such double intersections have been probably observed in ARPES data
of BSCCO \cite{1,2}.
Direct comparison with experiment would require to take into account the
bilayer structure of BSCCO and the corresponding band splitting.

Polarisational anisotropy of the FS directly reflects the existence of
the spin currents $J_{\ur}=-J_{\dr}$ in the spiral state.
According to \cite{28} it might be a reason of the
time reversal symmetry breaking observed in
the UD BSCCO in dichroism of the ARPES signal \cite{12}.
In the alternative hypothesis \cite{12,29} this effect is explained
by the specific alignment of the circular micro currents.

A direct observation of the polarisational anisotropy of the FS needs
the selective over the spin polarisation measurements of the photoemission
intensity. Such selective measurements of the total photoemission
have been achieved in the so called "spin-orbit" photoemission \cite{30}.
In principle, similar selectivity is possible in ARPES also.
It is important to continue the study of the time reversal symmetry breaking,
in particular, to clarify, whether this effect and accompanying currents
have the surface or bulk character.
Yet we cannot answer the question: if the ground state of the BSCCO
might be presented as a set of quasi-static domains with the spiral
structure and corresponding systems of spin currents.

In conclusion, the MF treatment of the normal state of the $t-t'-U$
Hubbard model shows that the FS topology and the PG anisotropy
depends on sign of $t'$, on the type (e- or h- ) of the doping
and of the spin structure.  For the hole-doped models the homogeneous
AF MF state occurs to be unstable with respect to formation of the stripe
phase and the spiral spin structure. However,  in electron-doped models
the lowest energy refers to the homogeneous AF solution.
This corresponds to a wide doping range of an existence of the local
AF order in the n-type cuprates NCCO, PCCO and to
commensurate peak at $Q=(\pi,\pi)$ observed in neutron scattering in
these cuprates.
In accordance with the ARPES data for the LSCO, the stripe phase
is characterized by the quasi-1D  behavior of the FS in the vicinity of
the "hot points" $(\pm\pi,0)$, by the pseudogap opening and by
suppression of the spectral weight in the diagonal direction
and in the direction parallel to stripes. Such anisotropy of FS and PG
does not conform with the d- symmetry of the SC order. For the spiral
spin structure the polarisation anisotropy of FS is revealed which
means that different segments of the FS correspond to different spin
electron polarisations.

Work is supported by the Russian Foundation of Basic Research under Grant
No.03-03-32141. Authors are greatfull to
V.Ya.Krivnov for stimulating discussions.

\newpage
Подписи к рисункам.

Fig.1.

In the top fig. the doping dependencies of the mean energy (per one site)
for the hole doped $t-t'-U$ model with parameters $U=4.0$, $t'=0.3$
for different spin structures are shown: for the paramagnet state (curve PM),
for the homogeneous AF state (curve AF), for the stripe phase with a
period of 8a along the x-axis (dashed line 1), for the spiral states with
vectors
$Q=\pi(\eta,1)$ (curve 2) and  $Q=\pi(\eta,\eta)$ (curve 3) at $\eta=0.8$.
The bottom graphics- the energies of the same structures (with the same
notations) for the electron-doped model. For convenience the common function
$F(\d)=U(n^2-1)$ is subtructed from all energies.

Fig.2.

a. A map of the photoemission intensity $I(k,\o=0)$ for
the antiphase AF stripes parallel to the $y$ axis,
having a width 4a and the bond-centered domain walls.
b. The same, but averaged over two stripe phases
of $x$- and $y$- orientation. Parameters of model are $U=6$, $t'=0.1$.

Fig.3.

a. The eigenenergies $E_{\l}(k)$ of the MF problem as functions of
the  quasi-momentum running along the contour
$Y(-\pi,\pi)$-$M_y(0,\pi)$-$Y(\pi,\pi)$-$M_x(\pi,0)$-$Y(\pi,-\pi)$
for the stripe structure with period 8a along the x-axis.
b. The map of the intensity $I(k,\o)$ for $k$,
running along the same contour. The map displays the same levels
$E_{\l}(k)$, but with a spectral weight, determined by the structure
of band states. Parameters of the model are the same, as in Fig.2.

Fig.4.

a. The map of the photoemission intensity $I_{\ur}(k,\o=0)$ for the electrons
with a spin polarisatin $\s=\ur$ for the spiral structure with
$Q=\pi(0.8,1)$. b. The same but averaged over two spin
polarisations $\s=\ur,\dr$ and over two structures with $Q=(0.8\pi,\pi)$
and $(\pi,0.8\pi)$. The model parameters are $U=6$, t'=0.1.

\newpage

\end{document}